# NX-CGRA: A Programmable Hardware Accelerator for Core Transformer Algorithms on Edge Devices

Rohit Prasad

Université Paris-Saclay, CEA, List, F-91120, Palaiseau, France, *rohit.prasad*@cea.fr

*Abstract*—The increasing diversity and complexity of transformer workloads at the edge present significant challenges in balancing performance, energy efficiency, and architectural flexibility. This paper introduces NX-CGRA, a programmable hardware accelerator designed to support a range of transformer inference algorithms, including both linear and non-linear functions. Unlike fixed-function accelerators optimized for narrow use cases, NX-CGRA employs a coarse-grained reconfigurable array (CGRA) architecture with software-driven programmability, enabling efficient execution across varied kernel patterns. The architecture is evaluated using representative benchmarks derived from real-world transformer models, demonstrating high overall efficiency and favorable energy-area tradeoffs across different classes of operations. These results indicate the potential of NX-CGRA as a scalable and adaptable hardware solution for edge transformer deployment under constrained power and silicon budgets.

## I. Introduction

Transformer-based models have become foundational in modern deep learning, offering superior accuracy and generalization across domains such as text, vision, and audio [16]. Their adoption is accelerating in edge-centric applications, including autonomous systems, healthcare monitoring, and IoT analytics, due to their capacity for efficient context modeling and parallel computation. However, the deployment of transformer inference on resource-constrained edge devices presents substantial architectural challenges due to the computational intensity and memory footprint of both linear and non-linear components of these models [7].

A significant portion of computation in transformers is derived from linear algebra operations, particularly matrix multiplications used in multi-head attention and feed-forward layers [7]. These operations become increasingly demanding as input dimensions grow, especially under low-latency constraints. In parallel, non-linear functions such as softmax, GELU, and layer normalization, while individually less compute-bound, contribute significantly to inference latency and energy consumption due to their complex mathematical formulations and hardware-unfriendly characteristics [7]. For example, softmax operations can constitute a considerable portion of the total inference time in large-scale models, such as decoder-only transformers optimized for text generation [31].

Conventional hardware solutions, including two-dimensional systolic arrays and SIMD accelerators, tend to be optimized for square or regular-shaped matrix workloads, limiting their utilization efficiency under the irregular and sparse dimensions commonly encountered in real-world transformer inputs [7]. General-purpose GPUs, although capable of high parallelism, often incur suboptimal latency and energy overheads for edge inference, particularly when executing non-offloaded control-heavy operations [20]. Moreover, the control flow complexity and workload diversity of transformer models hinder sustained processing element (PE) utilization, leading to increased idle cycles and underutilized compute capacity [6].

In response, recent research has proposed a variety of specialized accelerators to improve efficiency, including configurable datapath engines [21], analog-digital hybrid SoCs [9], and quantization-aware softmax accelerators [8]. These efforts explore tradeoffs between flexibility, performance, and energy efficiency. Nonetheless, many of these architectures remain fixed-function or optimized for a narrow subset of operations, limiting their generality across evolving AI workloads. The lack of architectural flexibility becomes particularly critical as transformer models continue to diversify in structure and deployment requirements.

This paper introduces NX-CGRA, a programmable hardware accelerator designed to bridge the gap between specialization and flexibility in edge-oriented transformer inference. The proposed architecture is a software-programmable CGRA architecture that supports a wide range of kernel types, including both compute-bound linear operations and memory-sensitive non-linear functions. Unlike prior fixed-function designs, NX-CGRA enables software-defined mapping of diverse kernels, promoting high PE utilization and efficient data movement under strict power and area budgets. Through gate-level evaluation of real-world core transformer algorithms, this work demonstrates the architectural potential of NX-CGRA in delivering scalable and efficient AI inference for the edge.

The rest of the paper is organized as follows: section II discusses background and related work, section III gives an overview of NX-CGRA, section IV discusses the experimental results, and finally section V provides a conclusion.

## II. Background and Related Work

This section provides an overview of transformers, the computational challenges they present for edge deployment,

TABLE I: Comparative Overview of Transformer Accelerators

| Accelerator | Linear | Non-Linear | Program. | Target Domain |
|---|---|---|---|---|
| SIGMA [20] | Yes | No | Limited | Sparse GEMM |
| CONNA [19] | Yes | No | Moderate | General Matrix Ops |
| Gemmini [5] | Yes | No | High | General DNN Inference |
| DIANA [9] | Yes | Partial | Moderate | Hybrid Analog-Digital |
| RBE [1] | Yes | No | Moderate | DSP and TinyML |
| RedMulE [26] | Yes | No | Low | TinyML / Low Power |
| OpenGEMM [36] | Yes | No | High | Edge AI / RISC-V SoC |
| OPTIMUS [18] | Yes | Yes | Limited | Transformer Decoding |
| SpAtten [31] | Yes | Yes | Limited | NLP Transformers |
| ELSA [8] | No | Yes | Limited | Attention Pruning |
| Wang et al. [33] | Yes | Yes | Moderate | Global Attention Opt. |
| Keller et al. [13] | Yes | Partial | Moderate | Mixed-Precision Attn. |
| ITA [10] | Yes | No | Moderate | Quantized Transformers |
| TranCIM [27] | Yes | Yes | Limited | CIM-based Transformers |
| SoftAct [4] | No | Yes | Low | Softmax Acceleration |
| PICACHU [21] | Yes | Yes | High | Non-linear LLM Ops |
| **NX-CGRA** | **Yes** | **Yes** | **High** | **General Transformer Inference** |

and a review of relevant accelerator architectures categorized by key optimization strategies.

*Transformers and Their Computational Challenges*: Deep learning algorithms, notably convolutional neural networks (CNNs), have significantly advanced fields such as computer vision [32], speech recognition [30], and natural language processing [2]. More recently, transformer-based architectures have emerged as the dominant paradigm, surpassing traditional recurrent and convolutional models across numerous AI domains [2], [23], [29]. The core innovation of transformers lies in the self-attention mechanism, which enables the model to capture global dependencies across input sequences by computing pairwise token interactions [29]. This results in superior performance in tasks requiring long-range context understanding.

Despite these advantages, transformers are computationally demanding and memory-intensive, posing major obstacles for their deployment on resource-constrained edge platforms. A typical transformer layer comprises attention mechanisms and fully connected layers, involving operations such as matrix multiplication (e.g., $QK^T$ and $PV$) that scale quadratically with the input sequence length [7]. This scaling results in substantial latency and energy costs. Non-linear operations, particularly softmax, GELU, and layer normalization, further exacerbate hardware implementation complexity due to their mathematical properties and high precision requirements [14]. For example, softmax computation has been shown to significantly contribute to overall inference delay in larger transformer models, highlighting the need for efficient implementations, even in scaled-down edge variants [31].

Edge devices, characterized by strict power budgets (averaging a few mW, with peaks of a hundred mW), and requiring low latency, necessitate specialized hardware acceleration to handle these computationally intensive workloads effectively [34]. General Matrix-Matrix Multiplication (GEMM) operations are the core computational kernels within most deep learning workloads, appearing frequently in both forward and backward passes [20]. While GPUs are advantageous for accelerating large amounts of computation through parallel processing, they are often not capable of achieving low latency and suffer from poor energy efficiency, especially for matrix multiplications of non-squared matrices with various shapes and dimensions that are common in transformer models adapted for edge use [6]. Therefore, numerous studies have attempted to design hardware accelerators with improved low latency and energy efficiency for neural networks [1], [5], [9], [19], [20], [26], [36].

*Linear Function Acceleration*: Many accelerators employed systolic arrays and SIMD architectures to accelerate dense matrix computations. Examples include Google's TPU [12] and academic efforts like Gemmini [5], CONNA [19], and OpenGeMM [36]. While efficient for regular matrix shapes, these designs typically underperform on the non-square and sparse matrix operations common in transformer workloads. Recent developments aim to enhance flexibility and configurability. CONNA introduces reconfigurable datapaths to maintain high utilization. Gemmini and OpenGeMM provide hardware generators and software stacks that support parameterized designs. MARSELLUS [1] integrates DSP cores with reconfigurable engines to balance flexibility and power scalability within the RISC-V SoC.

*Sparsity Exploitation*: Sparsity is a common characteristic in transformer attention maps, as many token pairs exhibit weak correlations. Several accelerators leverage this property. OPTIMUS [18] skips redundant operations via a specialized sparse matrix format. ELSA [8] filters irrelevant query-key pairs using approximations, reducing computation. SpAtten [31] employs pruning strategies and a top-k engine to eliminate low-importance tokens and heads. Wang et al. [33] propose a processor with speculative execution and dynamic scheduling to efficiently handle sparsity. SIGMA [20] provides a GEMM engine capable of sustaining high utilization despite irregular sparsity patterns using a flexible datapath and accumulation network.

*Quantization and Mixed-Precision Computing*: Reducing arithmetic precision is an effective strategy for improving energy efficiency. Architectures such as ITA [10] employ integer-only softmax implementations and 8-bit quantization. Keller et al. [13] demonstrate the effectiveness of per-vector 4-bit quantization and approximate softmax for transformers. RedMulE [26] supports FP16 and FP8 formats, enabling efficient low-power training and inference in TinyML-class devices. These designs emphasize minimal data movement and precision-aware computation.

*Compute-In-Memory (CIM)*: CIM architectures aim to reduce data movement by integrating computation within memory arrays. TranCIM [27] utilizes bitline-transpose CIM macros with a reconfigurable streaming network for sparse transformer execution. DIANA [9] integrates analog in-memory computation with digital reconfigurable cores to balance efficiency and accuracy in a heterogeneous SoC design. These approaches are particularly effective for reducing external memory access bottlenecks.

*Non-linear Function Acceleration*: Non-linear functions represent a critical bottleneck in transformer inference. Cross-Road Accel [14] introduces a specialized softmax architecture using base-2 and one-hot approximations. SoftAct [4] combines hardware-efficient softmax and activation units with algorithm-level pruning and detection to optimize power and area. PICACHU [21] targets activation-heavy operations in compact models using a CGRA and compiler co-design to support diverse non-linearities. These accelerators aim to maintain precision while minimizing the overhead of non-linear computation.

To provide a holistic view of recent hardware efforts targeting transformer inference at the edge, a comparative summary of representative accelerator designs is presented in Table I. The table provides a high-level categorization of hardware accelerators based on their ability to execute linear and non-linear workloads, programmability, and their targeted functionality within transformer models.

Despite substantial progress across these architectural classes, achieving an optimal balance of flexibility, latency, and energy efficiency remains an open challenge. The diversity and dynamism of modern transformer workloads on edge devices necessitate a new class of accelerators that are both programmable and efficient, motivating the design principles of NX-CGRA.

## III. NX-CGRA

NX-CGRA (see Figure 1 (a)) is a programmable hardware accelerator designed to support heterogeneous transformer inference workloads at the edge. Its microarchitecture is organized as a CGRA tailored for the efficient execution of both linear and non-linear computation kernels through a multiple instruction, multiple data (MIMD)-style execution model.

### A. Array Topology and Execution Model

The core of NX-CGRA consists of a two-dimensional heterogeneous array of PEs and Memory-Operation Blocks (MOBs) interconnected via a switchless mesh torus interconnect network (see Figure 1 (b)). Each core executes a microcode of the target application, statically scheduled during compile time. This design choice enables flexible and fine-grained scheduling of operations, particularly valuable for irregular workloads and non-uniform computation graphs typical of transformer models.

To facilitate MIMD execution, the array supports distributed instruction decoding and local register files. Each core is equipped with its own lightweight instruction buffer and controller to execute its assigned micro-operations.

### B. Heterogeneous Core Types: PEs and MOBs

NX-CGRA implements a custom VLIW-based ISA and comprises two distinct types of cores:

*1) Processing Elements (PEs):* PEs are arithmetic-capable cores optimized for executing the computational parts of the target application, and each PE, as shown in Figure 1(c), contains:

- Multi-precision integer-only modules supporting a range of operators across multiple data widths:
  - 8-bit operators include unsigned-unsigned multiply, 4× fused signed multiply-accumulate (MAC), saturate, and divide,
  - 16-bit unsigned-unsigned multiply operator, and
  - 32-bit operators include logic (bitwise operators, shift operators, comparison), branching (branch, conditional-branch), arithmetic (signed/unsigned add, subtract, signed-signed multiply, divide), and custom operations (sub-word masking, specialized data manipulation operators).
- Three register files (RFs): micro-code instruction RF, inline constant RF, and special temporary value RF supporting up to three operand reads with triple-read ports and single-port write access.
- A controller for instruction decoding, a core sleep unit, a clock gating unit, and routing interfaces for data exchange with neighboring PEs or MOBs, as well as for loading context data from the NX-Context Memory Subsystem via the NX-Memory Controller.

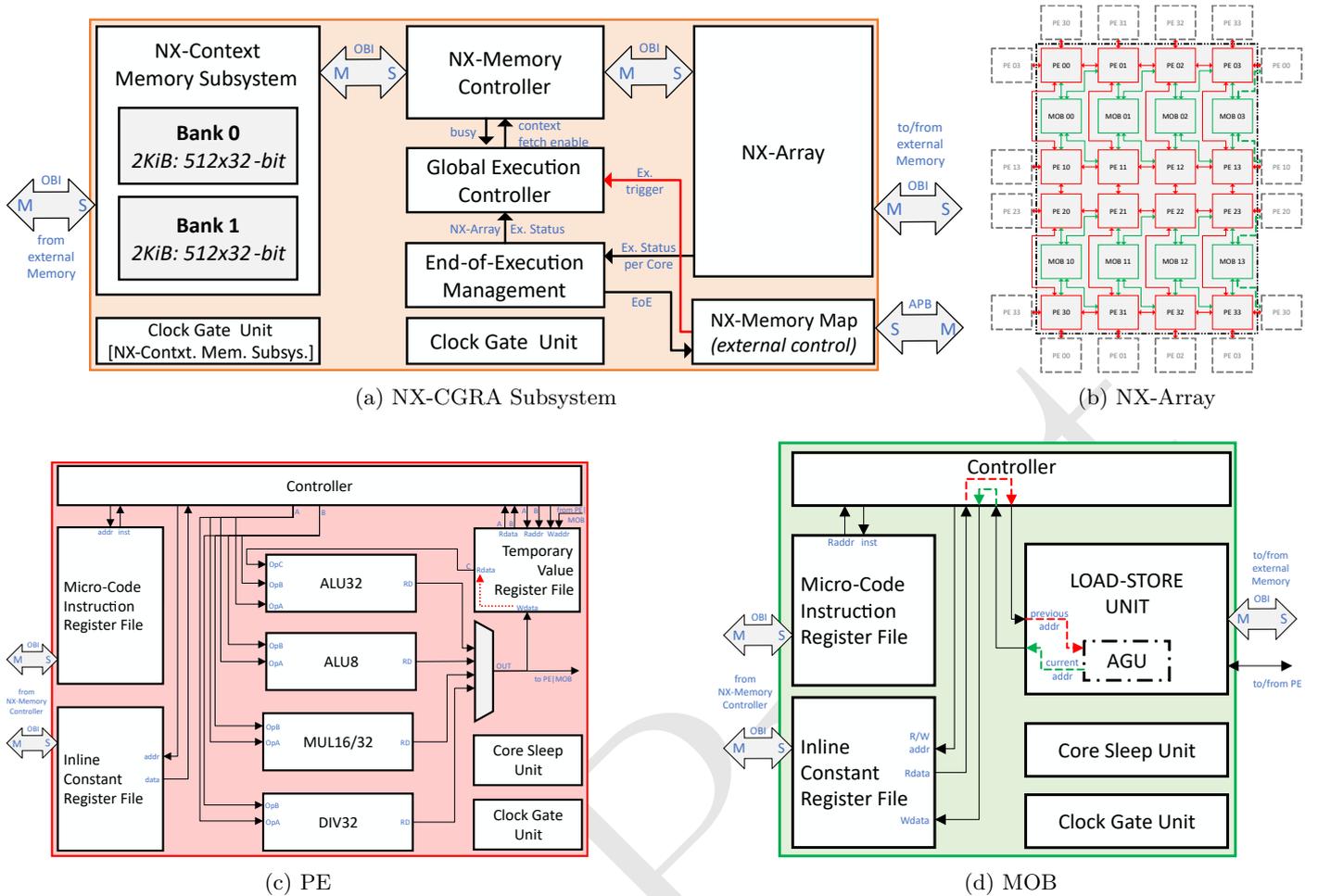

Fig. 1: Overview of the NX-CGRA subsystem and its components.
*Dashed lines in PE and MOB represent abstract data flow and do not indicate physical or direct hardware connections.*
*OBI=Open Bus Interface, APB=Advanced Peripheral Bus, M=Master, S=Slave, EoE=End-of-Execution, Ex.=Execution, PE=Processing Element, MOB=Memory-Operation Block, Rdata=Read Data, Waddr=Write Address, Wdata=Write Data, Op=Operand, RD=Destination Register, inst=Instruction, addr=Address, ALU32=32-bit ALU, ALU8=8-bit ALU, MUL16/32=16 & 32-bit Multiplication Unit, DIV32=32-bit Division Unit, AGU=Address Generation Unit*

*2) Memory Operation Blocks (MOBs):* MOBs are data-centric cores responsible for interfacing with external memories—implemented via master channels of the OBI protocol—performing memory address computations, and exchanging data with neighboring PEs (see Figure 1 (d)). Their architecture is similar to that of PEs; however, instead of ALUs, each MOB implements a Load-Store Unit (LSU) with an Address Generation Unit (AGU) to exclusively perform memory operations and branching decisions (branch and conditional branch). The primary function of MOBs is to facilitate spatial-temporal data locality and to mask the latency of data exchanges with external memories while the PEs perform computations. This way, MOBs create a separate channel by handling memory operations exclusively and effectively masking their latency during PE computation. The asymmetric design between PEs and MOBs enables a clear separation of compute and data movement responsibilities, optimizing pipeline utilization and interconnect usage.

### C. Data Movement, Synchronization, and Programmability

Data routing within the NX-CGRA is managed through statically scheduled MOVE operations, which are used to route data and enable data reuse. The compiler knows the lifetime of each variable at compile time. This information allows for optimized data routing, minimizing duplicate data fetching from external memory and thereby accelerating application execution on NX-CGRA.

The compiler generates micro-code for each core by scheduling and mapping the application onto the target NX-CGRA hardware. Control synchronization across cores is achieved through the use of JUMP and conditional JUMP (CJUMP) instructions, which serve as synchronization barriers. Outside these barriers, core executions are independent, supporting a partially distributed control model within the subsystem. Such support also enables

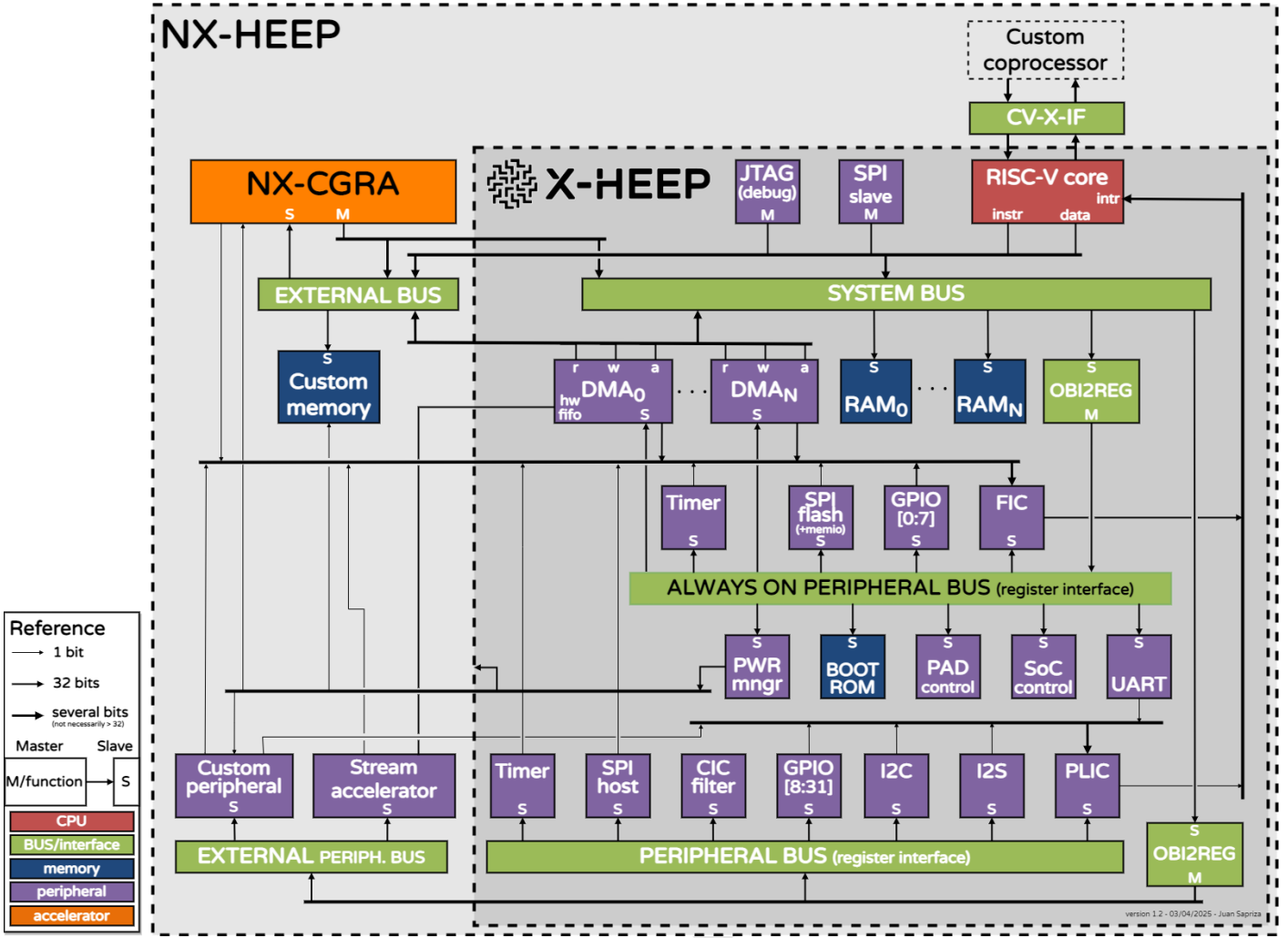

Fig. 2: NX-HEEP

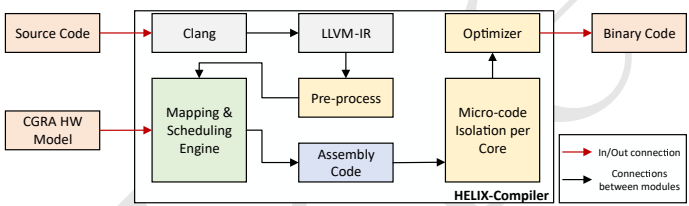

Fig. 3: NX-CGRA compilation toolchain

NX-CGRA to make branching decisions locally, allowing it to operate as a loosely coupled accelerator capable of running complete applications without relying on external host CPU control.

The compilation flow statically maps dataflow graphs derived from LLVM-IR onto the NX-CGRA fabric, producing static schedules for each core. While detailed compiler design and algorithms are outside the scope of this paper, this static scheduling framework enables efficient execution of workloads on NX-CGRA. An abstract overview of the compilation toolchain is shown in Figure 3.

### D. Subsystem Overview

Figure 1 (a) shows NX-CGRA subsystem overview. The subsystem consists of a $4 \times 6$ array of $4 \times 4$ PEs and $4 \times 2$ MOBs, a memory controller, a context memory subsystem, a global execution controller, an end-of-execution management, a memory map module, clock gating units, and multiple external interfaces. NX-CGRA is loosely coupled with a host CPU, enabling it to offload and execute complete applications independently. It implements a VLIW-based ISA and supports branching decisions. Data sharing between the NX-CGRA and host CPU is facilitated via a shared L1 memory composed of multiple banks connected through a low-latency logarithmic interconnect. Before execution, configuration data for each PE and MOB is stored in the context memory and distributed by the memory controller to their respective cores, ensuring full pre-configuration before application start.

### E. Integration into an Open-Source SoC

For seamless deployment, NX-CGRA has been integrated into an open-source SoC framework [17], maintaining com-

TABLE II: Benchmark Characteristics. †=int8, ‡=int16, §=int32

| Kernel | Input Size | % Kernel Composition | | | | | | |
|---|---|---|---|---|---|---|---|---|
| | | Tiny ViT [35] | Mobile BERT [25] | Tiny BERT [11] | Fast ViT [28] | Eff. FrmrV2 [15] | Whisper Tiny [22] | Distil BERT [24] |
| *conv*: 2D-Convolution | Img†  = $[3 \times 128 \times 128]$<br>Wgt† = $8*[3 \times 3 \times 3]$<br>Bias§ = $[8]$ | 25-30 | 0 | 0 | 60-65 | 55-60 | 0 | 0 |
| *gemm*: General Matrix Multiplication | A†=$[32 \times 64]$<br>B†=$[64 \times 32]$ | 45-55 | 60-70 | 60-70 | 15-20 | 20-25 | 65-70 | 65-70 |
| *gelu*: Activation | Input†=$[4 \times 16]$<br>Weight†=$[16]$<br>Bias§=$[16]$ | ∼5 | ∼5 | ∼5 | ∼5 | 5-8 | ∼5 | ∼5 |
| *norm*: Layer Normalization | Input†=$[64]$<br>Gamma†=$[8]$<br>Beta†=$[8]$ | ∼5 | 5-8 | 5-8 | 5-6 | 5-7 | 5-8 | 5-8 |
| *quant*: Quantization | Input‡=$[64]$<br>Scale§=$[1]$ | 0 | 0-5 | 0-5 | 0-5 | 0-5 | 0-5 | 0-5 |
| *sftmx*: SoftMax | QK_BUF†=$[32]$<br>ATTN_MASK§=$[32]$<br>BIAS§=$[32 \times 32]$ | ∼5 | ∼5 | ∼5 | 3-5 | 3-5 | ∼5 | ∼5 |

*% Kernel composition values are derived from architectural breakdowns and representative execution time distributions across layers, based on an analysis of the source code.*

patibility with standard edge processors through loose coupling to the host CPU. This integration utilizes a shared memory architecture to enable efficient data exchange between the NX-CGRA and the host processor (see Figure 2). A lightweight runtime system manages task offloading, synchronization, and power management on the NX-CGRA.

The NX-HEEP SoC also incorporates a Direct Memory Access (DMA) engine that supports asynchronous data transfers, minimizing CPU involvement during NX-CGRA operation. This approach reduces computational overhead and enhances energy efficiency.

## IV. EVALUATION

### A. Evaluation Methdology

To enable realistic evaluation, NX-CGRA is integrated into an open-source SoC environment, i.e., X-HEEP [17], and the performance metrics are obtained through gate-level simulations. While the X-HEEP documentation [3] provides general guidelines for customizing the platform, the modifications introduced here are specifically tailored to support NX-CGRA integration and the execution of edge transformer workloads. In particular, the L1 memory sub-system has been extended by placing additional $8 \times 32$ KiB interleaving memory banks, totaling 256 KiB of usable data space for the NX-CGRA. This design choice enables eight parallel memory accesses, aligned with the eight LSUs in NX-CGRA.

*1) Benchmarks:* A set of real-world transformer algorithms has been selected, representing typical workloads in compact, edge-oriented models. Table II summarizes the characteristics of each benchmark, including input sizes with corresponding data types (i.e., int8, int16, and int32), as well as their proportional contribution to transformer inference execution in these models.

The size of *sftmx* is too large to fit into the NX-CGRA. Consequently, this kernel has been divided into two parts. The intermediate values are stored in memory and reloaded to switch context data in the CGRAs.

In the case of *quant*, we selected the upper bound for the operator choice to account for situations where the data does not fit the operator input size. For example, if the input data is 16-bit and the operator is not available in the PE, we opted for 32-bit operators to carry out the operation. This was done to ensure that the accuracy of the result is not degraded.

*2) CAD Tools:* ASIC design flow was followed to conduct the experiments and record the performance metrics from NX-CGRA using the CAD tools listed below:

- Siemens Questasim (gate-level simulation)
- Synopsys Design Compiler (topographical synthesis)
- Synopsys PrimePower (power analysis)

### B. Experimental Results

*1) Implementation:* NX-CGRA is implemented in a 22nm FD-SOI technology, and the netlist is obtained by performing topographical synthesis (frequency = 200MHz, operating voltage = 0.8V, typical-typical analysis corner, and 25°C) using Synopsys Design Compiler.

Table V shows the total cell area breakdown of NX-CGRA sub-system, including a 4 KiB Context Memory,

TABLE III: Comparison of Micro-Architectures for Linear Transformer Functions

| Accelerator | SIGMA | CONNA | Gemmini | DIANA | RBE | RedMule | OpenGEMM | NX-CGRA |
|---|---|---|---|---|---|---|---|---|
| Publication | [20] | [19] | [5] | [9] | [1] | [26] | [36] | This Work |
| Year | 2020 | 2022 | 2021 | 2023 | 2024 | 2023 | 2025 | |
| Tech. ($nm$) | 28 | 65 | 16 | 22 | 22 | 22 | 16 | **22** |
| Area ($mm^2$) | 65.1 | 2.36 | 1.21 | 8.91 | 2.42 | 0.73/0.64 | 0.62 | **0.178** |
| Memory ($KiB$) | 6000 | 172 | 328 | 512 | 128 | 128 | 270 | **9.28/256** |
| Freq. ($MHz$) | 500 | 200 | 500 | 250 | 420 | 470/613 | 200 | **200** |
| Power ($mW$) | 22.33 | 83.55 | 312.41 | - | 12.8-123 | 37.6-193 | 43.8 | **1.513** |
| Voltage ($V$) | - | - | - | 0.8 | 0.5-0.8 | 0.65/0.8 | 0.675 | **0.8** |
| Precision | bfp16/fp32 | int4/16/32 | int8 | int2/4/8 | int2/8 | fp8/16 | int2/4/8 | **int8/16/32** |
| MACs | 16384 | 256 | 1024 | 256 | - | - | - | **64** |
| $GOPS$ | 16000 | 102.4 | 256 | 220 | 637 | 44.8-117 | 204 | **3.04** |
| $TOPS/W$ | 0.48 | 1.226 | 0.8195 | 4.1 | 12.4 | 0.506-1.666 | 4.68 | **2.01** |
| $TOPS/W/mm^2$ | 0.0073 | 0.52 | 0.6773 | 0.46 | 5.12 | 0.79/2.28 | 7.55 | **11.29** |

TABLE IV: Comparison of Micro-Architectures for Non-Linear Transformer Functions

| Accelerator | OPTIMUS | SpAtten | ELSA | Wang et al. | Keller et al. | ITA | TranCIM | SoftAct | NX-CGRA |
|---|---|---|---|---|---|---|---|---|---|
| Publication | [18] | [31] | [8] | [33] | [13] | [10] | [27] | [4] | This Work |
| Year | 2020 | 2021 | 2021 | 2022 | 2022 | 2023 | 2023 | 2024 | |
| Tech. ($nm$) | 28 | 40 | 40 | 28 | 5 | 22 | 28 | 28 | **22** |
| Area ($mm^2$) | 5.2 | 18.71 | 1.255 | 6.82 | 0.153 | 0.173 | 6.69 | 0.00287/0.00445 | **0.178** |
| Memory ($KiB$) | 1420 | 392 | 4.5 | 336 | 141 | 2.24 | - | - | **9.28/256** |
| Freq. ($MHz$) | 200 | 1000 | 1000 | 50-510 | 152-1760 | 500 | 80-240 | 1850/1560 | **200** |
| Power ($mW$) | 731.8 | 2600 | 969.36 | 12.06-272.8 | - | 60.5 | 27.04-118.21 | 3.46/5.76 | **1.617** |
| Voltage ($V$) | 1 | 1.1 | 1.1 | 0.56-1.1 | 0.46-1.05 | 0.8 | 0.6-1.0 | 0.9 | **0.8** |
| Precision | fixed16/32 | int8-16/fp32 | int8/fp16 | int12 | int4/8 | int8 | int8/16 | int16/32 | **int8/16/32** |
| MACs | 1024 | 1024 | 528 | 512 | 1024 | 1024 | - | - | **64** |
| $GOPS$ | 500.05 | 1610 | 1088 | 520-4070 | 3600/1800 | 1020 | 1480/370 | 1.85/1.56 | **1.1** |
| $TOPS/W$ | 0.68 | 0.62 | 1.12 | 1.91-27.56 | 95.6/39.1 | 16.9 | 20.5/5.1 | 0.54/0.27 | **0.68** |
| $TOPS/W/mm^2$ | 0.13 | 0.03 | 0.89 | 0.28-4.04 | 624.84/255.55 | 97.68 | 3.06/0.76 | 188.02/60.63 | **3.83** |

TABLE V: NX-CGRA total cell area ($\mu m^2$) breakdown (22$nm$ FD-SOI).

| | Component | Area ($\mu m^2$) | Area (%) |
|---|---|---|---|
| Memory Map | | 206 | 0.12 |
| Memory Controller | | 164 | 0.10 |
| Context Memory | 2 × 2 KiB SRAM macros | 13,327 | 7.49 |
| NX-Array | 16 PE, 8 MOB | 164,195 | 92.23 |
| *other* | | 107 | 0.06 |
| **NX-CGRA** | | 177,999 | 100.00 |

TABLE VI: Key Performance Metrics of selected kernels on NX-CGRA

| Kernel | MOPS | GOPS/$mm^2$ | TOPS/W | TOPS/W/$mm^2$ |
|---|---|---|---|---|
| conv | 1902 | 10.68 | 1.28 | 7.20 |
| gemm | 3040 | 17.08 | 2.01 | 11.29 |
| gelu | 636 | 3.57 | 0.39 | 2.21 |
| norm | 70 | 0.39 | 0.04 | 0.24 |
| quant | 255 | 1.43 | 0.16 | 0.89 |
| sftmx | 1102 | 6.19 | 0.68 | 3.83 |

a Memory Controller, a Memory Map, and 4 × 4 PE - 4 × 2 MOB array. Overall, the total cell area of NX-CGRA subsystem is only 0.178$mm^2$, which is insignificant when integrated into a more extensive system.

*2) Performance Metrics and Efficiency Analysis:* To assess the computational and energy efficiency of the NX-CGRA, several key performance metrics are considered across a representative set of inference kernels commonly found in transformer models. The $MOPS$ metric captures throughput per kernel, with observed values ranging from 636 $MOPS$ ($gelu$) & 1102 $MOPS$ ($sftmx$) for element-wise and activation functions to 1,902 $MOPS$ ($conv$) & 3,040 $MOPS$ ($gemm$) for compute-intensive kernels.

The $GOPS/mm^2$ metric quantifies spatial efficiency, revealing values in the range of 0.39 $GOPS/mm^2$ ($norm$) to 17.08 $GOPS/mm^2$ ($gemm$), depending on the degree of parallelism and on-chip data reuse within the micro-architecture. This metric is particularly relevant for edge computing scenarios, where area-constrained silicon implementations are prevalent.

Energy efficiency was measured using $TOPS/W$. Linear kernels demonstrated up to 2.01 $TOPS/W$ ($gemm$), while non-linear operations exhibited lower values around 0.39 $TOPS/W$ ($gelu$) & 0.68 $TOPS/W$ ($sftmx$), due to less regular compute patterns and increased control overhead. Combined energy-area efficiency reached a peak of 11.29 $TOPS/W/mm^2$ ($gemm$) in the most optimized kernel-to-microarchitecture mapping.

These figures establish a baseline for comparative evaluation against existing domain-specific accelerators in both linear and non-linear inference workloads. Detailed results for each kernel are summarized in Table VI.

*3) State of the Art Comparison:* To provide a comparative baseline, the performance of NX-CGRA is evaluated against a range of recently published domain-specific accelerators optimized for transformer inference on edge platforms. Table III and Table IV present this comparison across two categories of computational kernels: linear functions (e.g., GEMM and Convolution) and non-linear functions (e.g., Softmax and GELU).

Across both kernel categories, NX-CGRA demonstrates superior overall efficiency in terms of $TOPS/W/mm^2$ when compared individually with all listed accelerators targeting linear functions, and maintains competitive performance in non-linear workloads. Unlike fixed-function architectures, NX-CGRA is designed as a programmable and general-purpose accelerator, developed to address the growing demand for adaptable AI silicon that bridges the gap between specialization and generalization. Rather than relying on static kernel-specific hardware, the architecture employs a programmable set of compute and control primitives that are orchestrated to support diverse kernel behaviors efficiently. This design philosophy enables high PE utilization and localized data reuse across a broad spectrum of transformer algorithms, ensuring scalable efficiency amid the increasing heterogeneity of AI workloads.

## V. Conclusion

This work presents NX-CGRA, a programmable hardware accelerator designed to address the increasing heterogeneity of transformer inference workloads at the edge. The microarchitecture balances the efficiency of specialization with the flexibility of a CGRA, enabling support for a range of core transformer algorithms through software-driven programmability. Systematic benchmarking across linear and non-linear functions demonstrates that NX-CGRA achieves competitive energy-area efficiency (TOPS/W/mm$^2$) compared to existing accelerator designs. Unlike fixed-function architectures optimized for narrow workloads, the programmability of NX-CGRA enables efficient adaptation to diverse execution patterns, making it particularly suitable for transformer workloads operating under tight silicon and power budgets. Future work will focus on advanced compiler co-design and optimized scheduling techniques to enhance runtime adaptability across diverse workload profiles. Additionally, end-to-end performance evaluation and token-level execution analysis will be integrated to characterize system behavior further and improve execution efficiency and generalization.

### Acknowledgments

This research was partially funded by the dAIedge project (HORIZON-CL4-2022-HUMAN-02-02, Grant Agreement Number: 101120726).

### References

[1] F. Conti, G. Paulin, A. Garofalo, D. Rossi, A. Di Mauro, G. Rutishauser, G. Ottavi, M. Eggiman, H. Okuhara, and L. Benini. Marsellus: A heterogeneous RISC-V AI-IoT end-node SoC with 2–8 b DNN acceleration and 30 *IEEE Journal of Solid-State Circuits*, 59(1):128–142, 2024.

[2] J. Devlin, M.-W. Chang, K. Lee, and K. Toutanova. BERT: Pre-training of deep bidirectional transformers for language understanding, 2019.

[3] EPFL. extending X-HEEP. https://x-heep.readthedocs.io/en/latest/Extending/index.html, 2023. Accessed: 2025-06-13.

[4] Y. Fu, C. Zhou, T. Huang, E. Han, Y. He, and H. Jiao. SoftAct: A high-precision softmax architecture for transformers supporting nonlinear functions. *IEEE Transactions on Circuits and Systems for Video Technology*, 34(9):8912–8923, 2024.

[5] H. Genc, S. Kim, A. Amid, A. Haj-Ali, V. Iyer, P. Prakash, J. Zhao, D. Grubb, H. Liew, H. Mao, A. Ou, C. Schmidt, S. Steffl, J. Wright, I. Stoica, J. Ragan-Kelley, K. Asanovic, B. Nikolic, and Y. S. Shao. Gemmini: Enabling systematic deep-learning architecture evaluation via full-stack integration, 2021.

[6] S. Hadjis, F. Abuzaid, C. Zhang, and C. Ré. Caffe con troll: Shallow ideas to speed up deep learning, 2015.

[7] T. J. Ham, S. J. Jung, S. Kim, Y. H. Oh, Y. Park, Y. Song, J.-H. Park, S. Lee, K. Park, J. W. Lee, and D.-K. Jeong. A$^3$: Accelerating attention mechanisms in neural networks with approximation, 2020.

[8] T. J. Ham, Y. Lee, S. H. Seo, S. Kim, H. Choi, S. J. Jung, and J. W. Lee. ELSA: Hardware-software co-design for efficient, lightweight self-attention mechanism in neural networks. In *2021 ACM/IEEE 48th Annual International Symposium on Computer Architecture (ISCA)*, pages 692–705, 2021.

[9] P. Houshmand, G. M. Sarda, V. Jain, K. Ueyoshi, I. A. Papistas, M. Shi, Q. Zheng, D. Bhattacharjee, A. Mallik, P. Debacker, D. Verkest, and M. Verhelst. DIANA: An end-to-end hybrid digital and analog neural network soc for the edge. *IEEE Journal of Solid-State Circuits*, 58(1):203–215, 2023.

[10] G. Islamoglu, M. Scherer, G. Paulin, T. Fischer, V. J. Jung, A. Garofalo, and L. Benini. ITA: An energy-efficient attention and softmax accelerator for quantized transformers. In *2023 IEEE/ACM International Symposium on Low Power Electronics and Design (ISLPED)*, page 1–6. IEEE, Aug. 2023.

[11] X. Jiao, Y. Yin, L. Shang, X. Jiang, X. Chen, L. Li, F. Wang, and Q. Liu. TinyBERT: Distilling bert for natural language understanding, 2020.

[12] N. P. Jouppi, G. Kurian, S. Li, P. Ma, R. Nagarajan, L. Nai, N. Patil, S. Subramanian, A. Swing, B. Towles, C. Young, X. Zhou, Z. Zhou, and D. Patterson. TPU v4: An optically reconfigurable supercomputer for machine learning with hardware support for embeddings, 2023.

[13] B. Keller, R. Venkatesan, S. Dai, S. G. Tell, B. Zimmer, W. J. Dally, C. Thomas Gray, and B. Khailany. A 17–95.6 TOPS/W deep learning inference accelerator with per-vector scaled 4-bit quantization for transformers in 5nm. In *2022 IEEE Symposium on VLSI Technology and Circuits (VLSI Technology and Circuits)*, pages 16–17, 2022.

[14] J.-H. Kim, C.-H. Kim, S.-M. Rho, and K.-S. Chung. A low power attention and softmax accelerator for large language models inference. In *2024 IEEE International Conference on Consumer Electronics-Asia (ICCE-Asia)*, pages 1–4, 2024.

[15] Y. Li, J. Hu, Y. Wen, G. Evangelidis, K. Salahi, Y. Wang, S. Tulyakov, and J. Ren. Rethinking vision transformers for MobileNet size and speed. In *Proceedings of the IEEE international conference on computer vision*, 2023.

[16] T. Lin, Y. Wang, X. Liu, and X. Qiu. A survey of transformers, 2021.

[17] S. Machetti, P. D. Schiavone, T. C. Müller, M. Peón-Quirós, and D. Atienza. X-HEEP: An open-source, configurable and extendible risc-v microcontroller for the exploration of ultra-low-power edge accelerators, 2024.

[18] J. Park, H. Yoon, D. Ahn, J. Choi, and J.-J. Kim. OPTIMUS: Optimized matrix multiplication structure for transformer neural network accelerator. In I. Dhillon, D. Papailiopoulos, and V. Sze, editors, *Proceedings of Machine Learning and Systems*, volume 2, pages 363–378, 2020.

[19] S.-S. Park and K.-S. Chung. CONNA: Configurable matrix multiplication engine for neural network acceleration. *Electronics*, 11(15), 2022.

[20] E. Qin, A. Samajdar, H. Kwon, V. Nadella, S. Srinivasan, D. Das, B. Kaul, and T. Krishna. SIGMA: A sparse and irregular gemm accelerator with flexible interconnects for dnn training. In *2020 IEEE International Symposium on High Performance Computer Architecture (HPCA)*, pages 58–70, 2020.

[21] J. Qin, T. Xia, C. Tan, J. Zhang, and S. Q. Zhang. PICACHU: Plug-in CGRA handling upcoming nonlinear operations in llms. In *Proceedings of the 30th ACM International Conference on Architectural Support for Programming Languages and Operating*

*Systems, Volume 2*, ASPLOS '25, page 845–861, New York, NY, USA, 2025. Association for Computing Machinery.
[22] A. Radford, J. W. Kim, T. Xu, G. Brockman, C. McLeavey, and I. Sutskever. Robust speech recognition via large-scale weak supervision, 2022.
[23] A. Radford, J. Wu, R. Child, D. Luan, D. Amodei, and I. Sutskever. Language models are unsupervised multitask learners. *OpenAI*, 2019. Accessed: 2024-11-15.
[24] V. Sanh, L. Debut, J. Chaumond, and T. Wolf. DistilBERT, a distilled version of bert: smaller, faster, cheaper and lighter, 2020.
[25] Z. Sun, H. Yu, X. Song, R. Liu, Y. Yang, and D. Zhou. MobileBERT: a compact task-agnostic bert for resource-limited devices, 2020.
[26] Y. Tortorella, L. Bertaccini, L. Benini, D. Rossi, and F. Conti. RedMule: A mixed-precision matrix-matrix operation engine for flexible and energy-efficient on-chip linear algebra and tinyml training acceleration, 2023.
[27] F. Tu, Z. Wu, Y. Wang, L. Liang, L. Liu, Y. Ding, L. Liu, S. Wei, Y. Xie, and S. Yin. TranCIM: Full-digital bitline-transpose cim-based sparse transformer accelerator with pipeline/parallel reconfigurable modes. *IEEE Journal of Solid-State Circuits*, 58(6):1798–1809, 2023.
[28] P. K. A. Vasu, J. Gabriel, J. Zhu, O. Tuzel, and A. Ranjan. FastViT: A fast hybrid vision transformer using structural reparameterization. In *Proceedings of the IEEE/CVF International Conference on Computer Vision*, 2023.
[29] A. Vaswani, N. Shazeer, N. Parmar, J. Uszkoreit, L. Jones, A. N. Gomez, L. Kaiser, and I. Polosukhin. Attention is all you need, 2023.
[30] P. Verma and J. Berger. Audio transformers, 2025.
[31] H. Wang, Z. Zhang, and S. Han. SpAtten: Efficient sparse attention architecture with cascade token and head pruning. In *2021 IEEE International Symposium on High-Performance Computer Architecture (HPCA)*. IEEE, Feb. 2021.
[32] W. Wang, X. Yang, and J. Tang. Vision transformer with hybrid shifted windows for gastrointestinal endoscopy image classification. *IEEE Transactions on Circuits and Systems for Video Technology*, 33(9):4452–4461, 2023.
[33] Y. Wang, Y. Qin, D. Deng, J. Wei, Y. Zhou, Y. Fan, T. Chen, H. Sun, L. Liu, S. Wei, and S. Yin. A 28nm 27.5TOPS/W approximate-computing-based transformer processor with asymptotic sparsity speculating and out-of-order computing. In *2022 IEEE International Solid-State Circuits Conference (ISSCC)*, volume 65, pages 1–3, 2022.
[34] Y. Wang, Y. Qin, D. Deng, J. Wei, Y. Zhou, Y. Fan, T. Chen, H. Sun, L. Liu, S. Wei, and S. Yin. An energy-efficient transformer processor exploiting dynamic weak relevances in global attention. *IEEE Journal of Solid-State Circuits*, 58(1):227–242, 2023.
[35] K. Wu, J. Zhang, H. Peng, M. Liu, B. Xiao, J. Fu, and L. Yuan. TinyViT: Fast pretraining distillation for small vision transformers. In *European conference on computer vision (ECCV)*, 2022.
[36] X. Yi, R. Antonio, J. Dumoulin, J. Sun, J. Van Delm, G. Pereira Paim, and M. Verhelst. OpenGeMM: A highly-efficient gemm accelerator generator with lightweight risc-v control and tight memory coupling. In *Proceedings of the 30th Asia and South Pacific Design Automation Conference*, ASPDAC '25, page 1055–1061, New York, NY, USA, 2025. Association for Computing Machinery.